\documentclass[a4paper,superscriptaddress,showpacs,twocolumn,prl]{revtex4-1}
\usepackage{amsfonts}
\usepackage{amsmath}
\usepackage{amssymb}
\usepackage{graphicx}

\begin{document}

\title{Experimental analysis of the quantum complementarity principle}

\author{R. Auccaise}
\affiliation{Empresa Brasileira de Pesquisa Agropecu\'{a}ria, Rua Jardim Bot\^{a}nico 1024, 22460-000 Rio de Janeiro, Rio de Janeiro, Brazil}

\author{R. M. Serra}
\affiliation{Centro de Ci\^{e}ncias Naturais e Humanas, Universidade Federal do ABC, R.  Santa Ad\'{e}lia 166, 09210-170 Santo Andr\'{e}, S\~{a}o Paulo, Brazil}

\author{J. G. Filgueiras}
\affiliation{Centro Brasileiro de Pesquisas F\'{\i}sicas, Rua Dr. Xavier Sigaud 150, 22290-180 Rio de Janeiro, Rio de Janeiro, Brazil}

\author{R. S. Sarthour}
\affiliation{Centro Brasileiro de Pesquisas F\'{\i}sicas, Rua Dr. Xavier Sigaud 150, 22290-180 Rio de Janeiro, Rio de Janeiro, Brazil}

\author{I. S. Oliveira}
\affiliation{Centro Brasileiro de Pesquisas F\'{\i}sicas, Rua Dr. Xavier Sigaud 150, 22290-180 Rio de Janeiro, Rio de Janeiro, Brazil}

\author{L. C. C\'{e}leri}
\email{lucas.celeri@ufabc.edu.br}
\affiliation{Centro de Ci\^{e}ncias Naturais e Humanas, Universidade Federal do ABC, R.  Santa Ad\'{e}lia 166, 09210-170 Santo Andr\'{e}, S\~{a}o Paulo, Brazil}

\begin{abstract}
One of the milestones of quantum mechanics is Bohr's complementarity principle. It states that a single quantum can exhibit a particle-like \emph{or} a wave-like behaviour, but never both at the same time. These are mutually exclusive and complementary aspects of the quantum system. This means that we need distinct experimental arrangements in order to measure the particle or the wave nature of a physical system. One of the most known representations of this principle is the single-photon Mach-Zehnder interferometer. When the interferometer is closed an interference pattern is observed (wave aspect of the quantum) while if it is open, the quantum behaves like a particle. Here, using a molecular quantum information processor and employing nuclear magnetic resonant (NMR) techniques, we analyze the quantum version of this principle by means of an interferometer that is in a quantum superposition of being closed and open, and confirm that we can indeed measure both aspects of the system with the same experimental apparatus. More specifically, we observe with a single apparatus the interference between the particle and the wave aspects of a quantum system.
\end{abstract}

\maketitle


One of the most striking departure from the classical lines of thought is the double-slit experiment with a single quantum (from here now named qubit for simplicity). This experiment, which is an example of Bohr's complementarity principle, tells us that we have to choose either to observe interference fringes (wave-like behaviour) or to know which path has been taken by the qubit (particle-like behaviour). This fact, that has been experimentally verified in many different contexts \cite{DSE}, means that these two knowledges (wave-like and particle-like behaviour) are mutually exclusive. 

A possible realization of this experiment is the single-qubit Mach-Zehnder interferometer, schematically shown in Fig. 1. After crossing the first beam splitter, $BS_{1}$, the qubit is in a coherent superposition state of been taken both paths $a$ and $b$ at the same time. The second beam splitter, $BS_{2}$, if present, recombines the qubit paths that is, then, detected by $D_{a}$ or $D_{b}$. If we perform this experiment by varying the phase shift $\theta$ between the two paths, the result will be an interference pattern in the probability of the qubit detection by $D_{a}$ or $D_{b}$, indicating that the it behaves like a wave (for $\alpha = \pi$). However, if we remove $BS_{2}$ ($\alpha = 0$), the interference between both paths disappears and the particle character of the qubit is observed.

\begin{figure}[h]
\begin{center}
\includegraphics[scale=0.63]{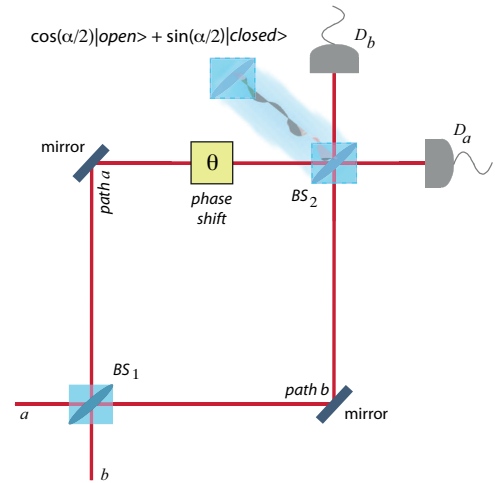}
\caption{(Colour Online) Schematic diagram of the Mach-Zehnder interferometer with a \textit{quantum} beam splitter $BS_{2}$. The interferometer is closed for $\alpha = \pi$ and open for $\alpha = 0$. For any other value, $0<\alpha<\pi$, the interferometer is in a coherent superposition of being closed and open. In the classical version of the complementarity experiment the beam splitter $BS_{2}$ has only two possible states, i.e., present or absent.}
\end{center}
\end{figure}

An important feature of classical complementarity experiment is the fact that, we can only say something about the behaviour of the system (particle or wave) \emph{after} the measurement has been carried out. We then must choose beforehand what phenomenon we want to observe. This means that the two experimental arrangements are complementary. This characteristics, which is the essence of Bohr's principle, led Wheeler to formulate his delayed-choice \textit{gedanken} experiment \cite{Wheeler}. Wheeler speculated whether the qubit could ``know'', before entering in the interferometer, what information we choose to observe and then behaves accordingly? To answer this question, Wheeler propose to make this choice only \emph{after} the qubit had crossed the first beam splitter. This experiment was recently performed in an optical setting and the complementarity principle was elegantly confirmed \cite{DCC}, showing that there is no difference between the normal and the delayed versions of the experiment. In other words, is only when we choose between particle or wave aspects that the reality emerges. It is worthwhile to note that this is a classical version of the experiment, in the sense that the interferometer has only two states, it is open \emph{or} closed. 

Here, using a nuclear magnetic resonant (NMR) setup, we experimentally study the quantum version of the delayed-choice experiment, recently proposed by Ionicioiu and Terno \cite{Terno}, in which the analogous of the second beam splitter of Fig. 1 is in a coherent superposition of being present and absent. In other words, \emph{the interferometer is in a quantum superposition of being closed and open}, which leads the qubit to be in a superposition of particle and wave. The obtained results agree with the theory and show that, contrary to the original statement of Bohr's principle, we can indeed use the same experimental arrangement to measure both aspects of the quantum system, i.e., wave and particle. This is accomplished due to the quantum nature of the controlling device, the second beam splitter in this case (as depicted in Fig. 1). 

The idea is as follows: The classical delayed choice experiment can be implemented by means of a quantum random number generator (QRNG). This QRNG is just an ancilla qubit $A$ that is prepared in an equal superposition state and measured after the system qubit $S$ had crossed the first beam splitter. The outcome (0 or 1) of this measurement, which is obtained before $S$ reaches the second beam splitter, will determine if the interferometer is open or closed \cite{DCC}. In the quantum version, this classical control is substituted by a quantum one, \emph{before} the measurement of the ancilla \cite{Terno}. This means that we can choose if $S$ will behave like a particle or like a wave \emph{after} it has been detected. Both behaviours are obtained by correlating the ancilla $A$ and the qubit $S$ experimental data.

Let us consider the input state (before $BS_{1}$ of Fig. 1) of the qubit as $\vert 0\rangle_{S}$ (where $\vert 0\rangle_{S}$ labels path $b$ and $\vert 1\rangle_{S}$ labels path $a$). The transformation employed by the $BS_{2}$ is coherently controlled by an ancillary qubit in such way that, if the ancilla particle is in the state $\vert 0\rangle_{A}$ the $BS_{2}$ is absent (the interferometer is open) and  the final state is $\vert 0\rangle_{A}\vert\mbox{particle}\rangle_{S}$ with $\vert\mbox{particle}\rangle_{S} := \left(\vert 0\rangle_{S}+ e^{i\theta}\vert 1\rangle_{S}\right)/\sqrt{2}$, representing the particle-like behaviour of the photon $S$. On the other hand, if the state of the ancilla is $\vert 1\rangle _{A}$ the $BS_{2}$ is present (the interferometer is closed) and the final state is $\vert 1\rangle_{A}\vert \mbox{wave}\rangle_{S}$ with $\vert \mbox{wave}\rangle_{S} := e^{i\theta/2}\left(\cos\left(\theta/2\right)\vert 0\rangle_{S} - i\sin\left(\theta/2\right) \vert 1\rangle_{S}\right)$, accounting for the wave-like behaviour of $S$. The ancilla qubit can also be prepared in a general coherent superposition, $\cos(\alpha/2)\vert 0\rangle _{A} + \sin(\alpha/2)\vert 1\rangle _{A}$, in this case, after $BS_{2}$, we obtain the state \cite{Terno}:
\begin{equation}
\vert \psi \rangle= \cos\left(\frac{\alpha}{2}\right)\vert 0\rangle_{A}\vert\mbox{particle}\rangle_{S}+ \sin\left(\frac{\alpha}{2}\right)\vert 1\rangle_{A}\vert\mbox{wave}\rangle_{S}.
\label{init}
\end{equation}

This scenario is completely different from the partial information of complementary quantities discussed in Ref. \cite{Englert}, in which a weak interaction with the interferometer allows us to obtain an imperfect knowledge about these quantities. In Ref. \cite{Englert} it was presented an inequality that introduces an upper bound to the maximum information, which we can simultaneously obtain about both aspects, interference fringe and which path. Here we have complete knowledge about both properties by letting the ``qubit'' (single quantum) to be in a superposition of both behaviours, particle and wave at the same time \cite{Terno}. 

\begin{figure}[h]
\begin{center}
\includegraphics[scale=0.34]{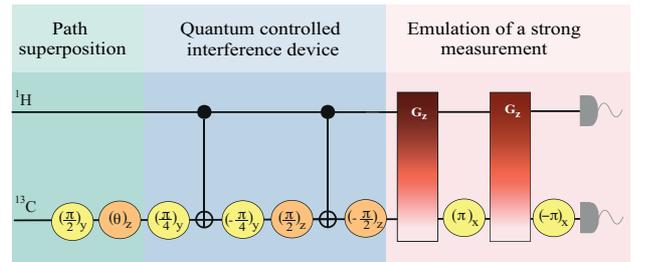}
\caption{(Colour Online). Sketch of the circuit for the quantum delayed-choice experiment. The first block represents the initial state preparation and employs the two interferometric paths. The second block performs a controlled interference between the two superposition paths encoded in the Carbon spin. A controlled-Hadamard gate is decomposed in four single qubit rotations and two CNOT gates. The CNOT gate is achieved by the radio-frequency pulse sequence (in time order) $\left(\frac{\pi}{2}\right)_{y}^{C}$-$U\left(\frac{1}{2J}\right)$-$\left(\frac{\pi}{2}\right)_{x}^{C}$-$\left(\frac{\pi}{2}\right)_{-y}^{C}$-$\left(\frac{\pi}{2}\right)_{-x}^{C}$-$\left(\frac{\pi}{2}\right)_{y}^{C}$-$\left(\frac{\pi}{2}\right)_{-y}^{H}$-$\left(\frac{\pi}{2}\right)_{x}^{H}$-$\left(\frac{\pi}{2}\right)_{y}^{H}$, where the super indices states for nucleus where the pulse is applied and $U\left(\frac{1}{2J}\right)$ represents a free evolution under the spin scalar coupling, $2\pi  J\mathbf{S}_{z}^{H}\mathbf{S}_{z}^{C}$, with $\mathbf{S}_{z}^{H(C)}=\hbar\sigma_{z}^{H(C)}/2$ being the spin angular momentum operator in the $z$-direction for $^{1}$H ($^{13}$C ) nuclei \cite{Ivan} and $J\approx 215.1$Hz in our experiment. Finally the third block emulates a strong measurement of the Hydrogen spin in the $\sigma_{z}$ eigenbasis by means of partial dephasing circuit. After this last block we perform a partial quantum state tomography of both qubits in order to obtain a signature of the wave-like or particle-like behaviour of the Carbon nuclear spin.}
\end{center}
\end{figure}

Our experiment was carried out in a liquid-state NMR spectroscopy with a two spin-$1/2$ sample of $^{13}$C-labelled CHCl$_{3}$. The sample was prepared by mixing 50 mg of 99 \% CHCl$_{3}$ in 0.7 ml of 99.9 \% Acetone-d6 in a 5 mm NMR tube. Both samples were provided by the Cambridge Isotope Laboratories - Inc. The experiments were performed at 25$^{\circ}$ C using a Varian 500 MHz Premium Shielded ($^{1}$H frequency) and a 5 mm double resonance probe-head equipped with a magnetic field gradient coil. 

Figure 2 shows the quantum circuit for the experimental procedure. The qubit encoded in the Hydrogen nuclear spin plays the role of the ancilla control for the interferometer device, while the qubit in the Carbon nuclear spin encompass the interferometer paths, i.e., the $^{13}$C magnetization aligned in the same direction of the spectrometer reference magnetic field represents the path $b$, while the magnetization aligned in the opposite direction represents path $a$, the states $ \vert 0\rangle_{S}$ and  $\vert 1\rangle_{S}$, respectively. The first block of Fig. 2 performs the initial state preparation and employs the superposition of the paths of the interferometer. By applying an appropriate combination of radio-frequency pulses and magnetic field gradients, we prepared the effective state $\vert 0 0 \rangle_{A,S}$. Although in liquid state NMR systems the whole ensemble cannot be prepared in a pure state, it is possible to prepare an analogous of such a state encoded in the so-called deviation matrix $\Delta\rho$. The state of the two-qubit spin ensemble can be represented in the high temperature expansion as $\rho=\mathbb{I}/4+\varepsilon\Delta\rho$, where $\varepsilon=\hbar\omega_{L}/4k_{B}T$ is the ratio between the magnetic and thermal energies \cite{NMRqbits,Chuang,Ivan} and $\mathbb{I}$ is the identity operator. The state of the spins system is highly mixed, however it can encompass coherent evolutions, interference \cite{Chuang,Ivan,Cory}, and also supporting general quantum correlations without entanglement \cite{PRA,PRL1,PRL2,Rev}. 

After such an initial preparation of the deviation matrix, a rotation $\left(\alpha\right)^{H}_{y}$ is applied (this notation means that a rotation of $\alpha$ in the $y$-direction is applied to the $^{1}H$ nucleus) in order to prepare the equivalent of the state $\cos(\alpha/2)\vert 0\rangle_{A} + \sin(\alpha/2)\vert 1\rangle_{A}$ of the ancilla qubit mapped into the Hydrogen magnetization state. A pseudo-Hadamard gate, $\left( \pi/2 \right)^{C}_{y}$, is applied to the $^{13}C$ nucleus in order to prepare the system's path superposition state $\left(\vert 0\rangle_{S} + \vert 1\rangle_{S}\right)/\sqrt{2}$. The phase shift between both interferometer arms is acquired by a rotation of $\theta$ in the $z$-direction applied to the Carbon nucleus. In the second block of Fig. 2 the quantum control of the interferometer (the analogous of the quantum $BS_{2}$ in Fig. 1) is implemented by means of a controlled-Hadamard gate \cite{Kumar} decomposed in four rotations and two CNOT gates as depicted in Fig. 2. The third block of the experimental procedure emulates a strong measurement of $\sigma_{z}$ eigenbasis on the ancilla qubit (${^{1}H}$) \cite{Teklemariam}. After this last block, the information about the particle-like or the wave-like nature of the single quantum system can be obtained from a partial quantum state tomography of the nuclear spins.

\begin{figure}[h]
\begin{center}
\includegraphics[scale=0.39]{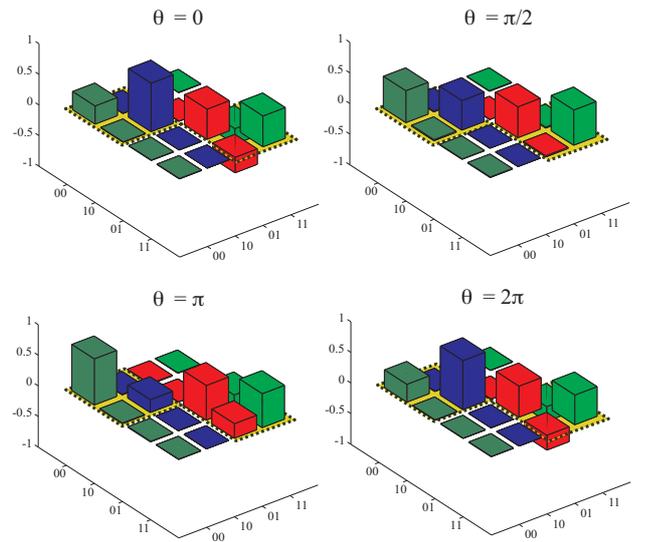}
\caption{(Colour Online). Real part of the final deviation matrix reconstructed by quantum state tomography for some values of the phase shift $\theta$ and for the parameter $\alpha=\pi/2$. The deviation matrix elements are displayed in the usual computational basis and the amplitudes are in arbitrary unities.}
\end{center}
\end{figure}

The emulation of a strong measurement is achieved as follows: The pulse sequence sketched in the third block of Fig. 2 leads to the dephasing of the ancilla spin (${^{1}H}$) by applying two identical linear gradient pulses. The coherence of the system spin ($^{13}C$) is protected by the $\left(\pi\right)^{^{13}C}_{x}$ pulse, which refocus its coherence. This procedure emulates a strong measurement in the $z$-direction, since the resulting density operator is exactly the same which we obtain in an ensemble of projective measures of $\sigma_{z}$ eigenbasis over the ancilla qubit. 

Employing such a strong measurement of the ancilla qubit in $z$-direction, we obtain the following deviation matrix
\begin{eqnarray}\nonumber
\Delta\rho_{Z} &\propto & \frac{\mathbb{I} + \sigma^{A}_{z}}{2}\cos\left(\frac{\alpha}{2}\right)^{2}\vert\mbox{particle}\rangle_{S}\langle\mbox{particle}\vert \\ 
&& + \frac{\mathbb{I} - \sigma^{A}_{z}}{2}\sin\left(\frac{\alpha}{2}\right)^{2}\vert\mbox{wave}\rangle_{S}\langle\mbox{wave}\vert.
\label{final}
\end{eqnarray}
The deviation matrix in Eq. (\ref{final}) is a block diagonal matrix with each block containing the wave-like or particle-like information about the behaviour of the system. Since we are dealing with an ensemble of molecules, we cannot perform a single projective measurement on the ancilla and then post-select the outcomes. However, we can emulate the non-unitary dynamics associated with the strong measurement processes by means of a partial dephasing circuit as already commented (see Fig. 2). The action of this operation is to completely dephase the coherences associated with the ancilla exactly in the same way they would be by a strong measurement of $\sigma_{z}$ eigenstates \cite{Teklemariam}. Following the dynamics of just one element of the deviation matrix (\ref{final}), we can also emulate the post-selection of a given result for the measurement of the ancilla qubit.  We illustrate such emulation of a projective measurement in Fig. 3 by showing the full quantum state tomography \cite{Ivan,Tomo} of the final deviation matrix for some values of the phase shift $\theta$. The block diagonal form of Eq. (\ref{final}) as well as the oscillations of the deviation matrix elements due to the path interference  can be clearly observed in Fig. 3. We note that it is not necessary to perform the full quantum state tomography to explore the wave-particle complementary aspects of a single spin. We can do it following the dynamics of just one element of the deviation matrix, i.e., that one which represents the input state of the interferometer. 

\begin{figure}[h]
\begin{center}
\includegraphics[scale=0.56]{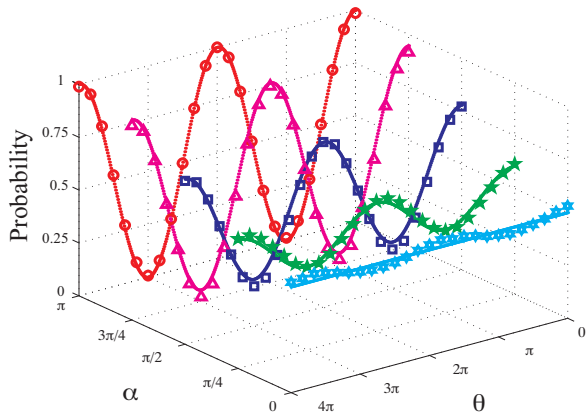}
\caption{(Colour Online). Probability of detect the initial state, $\mbox{Tr}\left(\Delta\rho\vert 1 0 \rangle_{A,S}\langle 1 0 \vert\right)$, emulating the strong measurement in the $\sigma_{z}$ eigenbasis, as function of $\theta$ for different values of $\alpha$. The dots represent the experimental points while the lines are cosine fittings of these experimental data.}
\end{center}
\end{figure}

In Fig. 4 we can observe a continuous transition, as function of the parameter $\alpha$, from the wave-like nature (the cosine oscillation as function of theta) to the particle-like nature (the flat probability as function of theta). These complementary aspects of the one single-quantum interferometer can be observed in the present experimental arrangement through measurements of nuclear magnetizations. 

In summary, we have successfully implemented the quantum version of Wheeler's delayed-choice experiment, theoretically proposed in \cite{Terno}. The obtained results are in very good agreement with the theoretical predictions and reveals some novel and important aspects of Bohr's complementarity principle. Some reinterpretation of this principle is in order. As we observed in the experimental results depicted in Fig. 4, it is possible to prepare and test a single quantum into a superposition of \emph{particle} and \emph{wave} aspects, which permits us to obtain both informations with the same experimental setting, since we have a device able to coherently control a path interference. This fact is in sharp contrast with the original statement of the complementarity principle. This forces us to disregard \textit{particle} and \textit{wave} aspects as realistic properties of the system \cite{Terno}. 

This fact could be analysed from the information theory perspective. The uncertainty about the system (in this case about two complementary variables) is not a property of the system, but it is a property of the experimental device employed to investigate the system's behaviour. The results theoretically presented in Ref. \cite{Terno} and experimentally demonstrated here suggest that a quantum device could obtain more information from the system than a classical one.

\emph{Note Added}: After completion of this manuscript, we became aware of an independent work \cite{QDCE} that have obtained similar conclusions by using a different technique, corroborating the present experiment. 

\begin{acknowledgments}
This work was supported by CNPq, FAPESP, FAPERJ, and the Brazilian National Institute for Science and Technology of Quantum Information (INCT-IQ). LCC specially thanks for the warm hospitality of the Brazilian Centre for Physics Research (CBPF) in Rio de Janeiro where the experiments were carried out. 
\end{acknowledgments}


\begin{thebibliography}{99}

\bibitem{DSE} A. Tonomura, J. Endo, T. Matsuda, T. Kawasaki, and H. Ezawa, Am. J. Phys. \textbf{57}, 117 (1989); J. Summhammer, G. Badurek, H. Rauch, U. Kischko, and A. Zeilinger, Phys. Rev. A \textbf{27}, 2523 (1983); O. Carnal and J. Mlynek, Phys. Rev. Lett. \textbf{66}, 2689 (1991); D. W. Keith, C. R. Ekstrom, Q. A. Turchette, and D. E. Pritchard, Phys. Rev. Lett. \textbf{66}, 2693 (1991); M. Arndt \emph{et al.}, Nature \textbf{401}, 680 (1999).

\bibitem{Wheeler} J. A. Wheeler, in \emph{Mathematical Foundations of Quantum Mechanics}, A. R. Marlow, ed. (Academic, New York, 1978).

\bibitem{DCC} V. Jacques, E. Wu, F. Grosshans, F. Treussart, P. Grangier, A. Aspect, and J.-F. Roch, Science \textbf{315}, 966 (2007).

\bibitem{Terno} R. Ionicioiu and D. R. Terno, Phys. Rev. Lett. \textbf{107}, 230406 (2011).

\bibitem{Englert} B.-G. Englert, Phys. Rev. Lett. \textbf{77}, 2154 (1996); V. Jacques, E. Wu, F. Grosshans, F. Treussart, P. Grangier, A. Aspect, and J.-F. Roch, Phys. Rev. Lett. \textbf{100}, 220402 (2008).

\bibitem{Cory} M. Pravia, E. Fortunato, Y. Weinstein, M. D. Price, G. Teklemariam, R. J. Nelson, Y. Sharf, S. Somaroo, C. H. Tseng, T. F. Havel, and D. G. Cory, Concepts Magn. Reson. \textbf{11}, 225 (1999).

\bibitem {NMRqbits} T. D. Ladd, F. Jelezko, R. Laflamme, Y. Nakamura, C. Monroe, and J. L. O'Brien, Nature \textbf{464}, 45 (2010).

\bibitem {Chuang} L. M. K.  Vandersypen and I. L. Chuang, Rev. Mod. Phys. \textbf{76}, 1037 (2004).

\bibitem {Ivan} I. S. Oliveira, T. J. Bonagamba, R. S. Sarthour, J. C. C.  Freitas, R. R. deAzevedo, \textit{NMR Quantum Information Processing} (Elsevier, Amsterdam, 2007).

\bibitem {PRA} D. O. Soares-Pinto, L. C. C\'{e}leri, R. Auccaise, F. F. Fanchini, E. R. deAzevedo, J. Maziero, T. J. Bonagamba, and R. M. Serra, Phys. Rev. A \textbf{81}, 062118 (2010).

\bibitem {PRL1} R. Auccaise, J. Maziero, L. C. C\'{e}leri, D. O. Soares-Pinto, E. R. deAzevedo, T. J. Bonagamba, R. S. Sarthour, I. S. Oliveira, and R. M. Serra, Phys. Rev. Lett. \textbf{107}, 070501 (2011).

\bibitem {PRL2} R. Auccaise, L. C. C\'{e}leri, D. O. Soares-Pinto, E. R. deAzevedo, J. Maziero, A. M. Souza, T. J. Bonagamba, R. S. Sarthour, I. S. Oliveira, and R. M. Serra, Phys. Rev. Lett. \textbf{107}, 140403 (2011).

\bibitem {Rev} For general information about quantum correlations without entanglement see: L. C. C\'{e}leri, J. Maziero, and R. M. Serra, Int. J. of Quantum Inf. \textbf{9}, 1837 (2011) and K. Modi, A. Brodutch, H. Cable, T. Paterek, and V. Vedral, arXiv:1112.6238 (2011).

\bibitem{Kumar} P. Kumar and S. R. Skinner. Phys. Rev. A \textbf{76}, 022335 (2007).

\bibitem{Teklemariam} G. Teklemariam, E. M. Fortunato, M. A. Pravia, T. F. Havel, and D. G. Cory. Phys. Rev. Lett. \textbf{86}, 5845 (2001).

\bibitem{Tomo} G. L. Long, H. Y. Yan, and Y. Sun, J. Opt. B: Quantum Semiclass. Opt. \textbf{3}, 376 (2001); J. Teles, E. R. deAzevedo, R. Auccaise, R. S. Sarthour, I. S. Oliveira, and T. J. Bonagamba, J. Chem. Phys. \textbf{126}, 154506 (2007).

\bibitem{QDCE} S. S. Roy, A. Shukla, and T. S. Mahesh, arXiv:1112.3524 (2011).

\end{thebibliography}
\end{document}